\documentclass[11pt]{article}
\usepackage{graphicx}
\usepackage{amsmath}
\usepackage{amssymb}
\usepackage{amsxtra}

\def\half{\frac{1}{2}}

\title{Unified description of quarks and leptons\\ in a multi-spinor field formalism}
\author{Ikuo S. Sogami\\ Yukawa Institute of Theoretical Physics, Kyoto University\\
Department of Physics, Kyoto Sangyo University\\
sogami@cc.kyoto-su.ac.jp\\}

\begin{document}
\maketitle

\begin{abstract}
Multi-spinor fields which behave as triple-tensor products of the Dirac spinors
and form reducible representations of the Lorentz group describe three families
of ordinary quarks and leptons in the visible sector and an additional family of
exotic dark quarks and leptons in the dark sector of the Universe. Apart from
the ordinary set of the gauge and Higgs fields in the visible sector, another set
of gauge and Higgs fields belonging to the dark sector are assumed to exist. 
Two sectors possess channels of communication through gravity and a bi-quadratic
interaction between the two types of Higgs fields. A candidate for the main
component of the dark matter is a stable dark hadron with spin 3/2, and
the upper limit of its mass is estimated to be 15.1 GeV/c$^2$.
\end{abstract}

\section{Introduction}
The Standard Model (SM) has been accepted as an almost unique effective scheme
for phenomenology of particle physics in the energy region around and lower than
the electroweak scale. Nevertheless it should be considered that the SM is still
in an incomplete stage, since its fermionic and Higgs parts are full of unknowns.
It is not yet possible to answer the question why quarks and leptons exist in
the modes of three families with the color and electroweak gauge symmetry
$G=SU_c(3)\times SU_L(2)\times U_Y(1)$, and we have not yet found definite rules
to determine their interactions with the Higgs field. It is also a crucial issue
to inquire whether the SM can be extended so as to accommodate the degrees of
freedom of dark matter.

To extend the SM to a more comprehensive theory which can elucidate its unknown
features, we introduce the algebra, called {\it triplet algebra},
consisting of triple-tensor products of the Dirac algebra and
construct a unified field theory with the multi-spinor field, called
{\it triplet field}, which behaves as triple-tensor products of the
four-component Dirac spinor~\cite{Sogami1,Sogami2}. The chiral triplet fields
forming reducible representations of the Lorentz group include the three families of
ordinary quarks and leptons and also an additional fourth family of exotic
quarks and leptons which are assumed to belong to the dark sector of the Universe.

The bosonic part of the theory consists of the ordinary gauge and Higgs fields
of the $G$ symmetry and also the dark gauge and Higgs fields of the new
$G_\star=SU_{c\star}(3)\times SU_R(2)\times U_{Y_\star}(1)$ symmetry. 
While the gauge fields of the $G$ symmetry interact with the ordinary quarks
and leptons of the three families, the gauge fields of the $G_\star$ symmetry
are presumed to interact exclusively with the quarks and leptons of the
fourth family in the dark sector. The gauge fields of the extra color symmetry
$SU_{c\star}(3)$ work to confine the dark quarks into dark hadrons. Apart from
the ordinary Higgs field $\varphi$ which breaks the electroweak symmetry
$G_{\rm EW}=SU_L(2)\times U_Y(1)$ at the scale $\Lambda$, another Higgs field
$\varphi_\star$ is assumed to exist to break the left-right twisted symmetry
$G_{\rm EW\star}=SU_R(2)\times U_{Y_\star}(1)$ at the scale $\Lambda_\star$
($\Lambda_\star > \Lambda$).

Our theory predicts existence of a stable dark hadron with spin 3/2 as a candidate
for the main component of the dark matter. From a heuristic argument, we estimate
the upper limit of its mass to be 15.1 GeV/c$^2$.

\section{Triplet field and triplet algebra}
To describe all fermionic species of the SM, we introduce the triplet field $\Psi(x)$
which behaves as triple-tensor products of the Dirac spinors as
\begin{equation}
     \Psi_{abc}\ \sim\ \psi_a\,\psi_b\,\psi_c 
\end{equation}
where $\psi$ is the four-component Dirac spinor. Operators acting on the triplet field
belong to the triplet algebra $A_T$ composed of the triple-tensor products of the Dirac algebra $A_\gamma = \langle\,\gamma_\mu\,\rangle=\{\,1, \gamma_\mu,\, \sigma_{\mu\nu},\, \gamma_5\gamma_\mu,\, \gamma_5 \,\}$ as follows: 
\begin{equation}
\begin{array}{lcl}
   A_T &=& \{\,p\otimes q\otimes r :\  p,\,q,\,r \in A_\gamma\,\}\\
   \noalign{\vskip 0.3cm}
       &=& \langle\, \gamma_\mu\otimes 1\otimes 1,\ 
       1\otimes \gamma_\mu \otimes 1,\  
       1\otimes 1\otimes \gamma_\mu\,\rangle .
\end{array}
\end{equation} 

The triplet algebra $A_T$ is too large for all its elements to acquire physical
meanings. To extract its subalgebras being suitable for physical description in
the SM energy region, we impose the criterion~\cite{Sogami1} that the subalgebra
bearing physical interpretation is closed and irreducible under the action of the
permutation group $S_3$ which works to exchange the order of $A_\gamma$ elements
in the tensor product. With this criterion, the triplet algebra can be decomposed
into three mutually commutative subalgebras, i.e., an external algebra defining
external properties of fermions and two internal subalgebras that have
the respective roles of prescribing family and color degrees of freedom.

The four elements
\begin{equation}
  \Gamma_\mu = \gamma_\mu\otimes\gamma_\mu\otimes\gamma_\mu \in A_T,
  \quad (\mu = 0,\,1,\,2,\,3)
\end{equation}
satisfy the anti-commutation relations
$\Gamma_\mu\Gamma_\nu + \Gamma_\nu\Gamma_\mu = 2\eta_{\mu\nu}I$ where 
$I = 1\otimes 1\otimes 1$. With them, let us construct an algebra $A_\Gamma$ by
\begin{equation}
  A_{\Gamma} = \langle\, \Gamma_{\mu}\, \rangle
         =\ \{\ I,\ \Gamma_\mu,\, \Sigma_{\mu\nu},\,
                    \Gamma_5\Gamma_\mu,\,\Gamma_5 \ \}
\end{equation}
where
$\Sigma_{\mu\nu}= -\frac{i}{2}(\Gamma_\mu\Gamma_\nu - \Gamma_\nu\Gamma_\mu)
=\sigma_{\mu\nu}\otimes \sigma_{\mu\nu}\otimes \sigma_{\mu\nu}$
and
$\Gamma_5 = -i\Gamma_0\Gamma_1\Gamma_2\Gamma_3 = \Gamma^5
        = \gamma_5 \otimes\gamma_5 \otimes \gamma_5$.
The algebra $A_\Gamma$ being isomorphic to the original Dirac algebra $A_\gamma$
fulfills the $S_3$ criterion and works to specify the external characteristics
of the triplet field. Namely, we postulate that the operators
$M_{\mu\nu} = \frac{1}{2}\Sigma_{\mu\nu}$ generate the Lorentz transformations
for the triplet field $\Psi(x)$ in the four dimensional Minkowski
spacetime $\{\,x^\mu\,\}$ where we exist as observers. The subscripts of
operators $\Gamma_\mu$ are related and contracted with the superscripts of
the spacetime coordinates $x^\mu$.

Under the proper Lorentz transformation $x^{\prime \mu} = \Omega^\mu{}_\nu x^\nu$, 
the triplet field and its adjoint field $\overline{\Psi}(x) = \Psi^\dagger(x)\Gamma_0$
are transformed as
\begin{equation}
 \Psi^\prime(x^\prime) = S(\Omega)\Psi(x),\quad 
 \overline{\Psi}^\prime(x^\prime) = \overline{\Psi}(x)S^{-1}(\Omega)
 \label{Lorentz1}
\end{equation}
where the transformation matrix is given by
\begin{equation}
   S(\Omega) = \exp\left(-\frac{i}{2}M_{\mu\nu}\omega^{\mu\nu}\right)
 \label{Lorentz2}
\end{equation}
with the angles $\omega^{\mu\nu}$ in the $\mu$--$\nu$ planes. The Lorentz invariant scalar
product is formed as
\begin{equation}
  \overline{\Psi}(x)\Psi(x)
  = \sum_{abc}\overline{\Psi}_{abc}(x)\Psi_{abc}(x).
  \label{Linvariant}
\end{equation}
For discrete transformations such as space inversion, time reversal and
the charge conjugation, the present scheme retains exactly the same structure as
the ordinary Dirac theory. The chirality operators are given by
\begin{equation}
    L = \frac{1}{2}(I - \Gamma_5),\quad
    R = \frac{1}{2}(I + \Gamma_5)\ \in\ A_{\Gamma}
\end{equation}
which are used to assemble algebras for electroweak symmetries.

Note that the Dirac algebra $A_\gamma$ possesses two $\mathfrak{su}(2)$ subalgebras
\begin{equation}
A_\sigma=\{\,\sigma_1=\gamma_0,\,\sigma_2=i\gamma_0\gamma_5,\,\sigma_3=\gamma_5\,\}\quad
\label{Asigma} 
\end{equation}
and
\begin{equation}
  A_\rho=\{\,\rho_1=i\gamma_2\gamma_3,\,\rho_2=i\gamma_3\gamma_1,\,\rho_3=i\gamma_1\gamma_2\,\}
\label{Arho}
\end{equation}
which are commutative and isomorphic with each other. By taking the triple-tensor
products of elements of the respective subalgebras $A_\sigma$ and $A_\rho$ in $A_T$,
we are able to construct two sets of commutative and isomorphic subalgebras with compositions {\lq}{\lq}$\mathfrak{su}(3)$ plus $\mathfrak{u}(1)${\rq}{\rq} which
satisfy the criterion of $S_3$ irreducibility. Those algebras are postulated to
have the roles to describe internal family and color degrees of freedom of the triplet fields.

\section{Algebra for extended family degrees of freedom}
From the elements of the algebra $A_\sigma=\{\,\sigma_1,\,\sigma_2,\,\sigma_3\,\}$ in Eq.(\ref{Asigma}),
we can make up eight elements of $A_T$ as follows:
\begin{equation}
\qquad \left\{\ 
  \begin{array}{l}
    \pi_1 = \frac{1}{2}\left(\sigma_1\otimes\sigma_1\otimes 1
          + \sigma_2\otimes\sigma_2\otimes 1\right),\\  
             \noalign{\vskip 0.3cm}
    \pi_2 = \frac{1}{2}\left(\sigma_1\otimes\sigma_2\otimes\sigma_3
          - \sigma_2\otimes\sigma_1\otimes\sigma_3\right),\\
             \noalign{\vskip 0.3cm}
    \pi_3 = \frac{1}{2}\left(1\otimes\sigma_3\otimes\sigma_3
          - \sigma_3\otimes 1\otimes\sigma_3\right),\\
          \noalign{\vskip 0.3cm}
    \pi_4 = \frac{1}{2}\left(\sigma_1\otimes 1\otimes\sigma_1
          + \sigma_2\otimes 1\otimes \sigma_2\right),\\
          \noalign{\vskip 0.3cm}  
    \pi_5 = \frac{1}{2}\left(\sigma_1\otimes\sigma_3\otimes\sigma_2
          - \sigma_2\otimes\sigma_3\otimes\sigma_1\right),\\
          \noalign{\vskip 0.3cm}
    \pi_6 = \frac{1}{2}\left(1\otimes\sigma_1\otimes\sigma_1
          + 1\otimes\sigma_2\otimes\sigma_2\right),\\
          \noalign{\vskip 0.3cm} 
    \pi_7 = \frac{1}{2}\left(\sigma_3\otimes\sigma_1\otimes\sigma_2
          - \sigma_3\otimes\sigma_2\otimes\sigma_1\right),\\
          \noalign{\vskip 0.3cm}
    \pi_8 = \frac{1}{2\sqrt{3}}\left(1\otimes\sigma_3\otimes\sigma_3
          + \sigma_3\otimes 1\otimes\sigma_3 
          -2\sigma_3\otimes\sigma_3\otimes 1\right) .
   \label{pidef}
\end{array}
\right.
\end{equation}
which are proved to obey the commutation and anti-commutation relations of
the Lie algebra $\mathfrak{su}(3)$ as 
\begin{equation}
        [\,\pi_j,\ \pi_k\,] = 2f^{(3)}_{jkl}\pi_l,\quad 
       \{\,\pi_j,\ \pi_k\,\} = \frac{4}{3}\delta_{jk}\Pi_{(v)}
        + 2d^{(3)}_{jkl}\pi_l
        \label{picom}
\end{equation}
where 
\begin{equation}
 \Pi_{(v)} = \frac{1}{4}\left(3I - 1\otimes1\otimes\sigma_3\otimes\sigma_3
 - 1\otimes\sigma_3\otimes 1\otimes\sigma_3
 -1\otimes\sigma_3\otimes\sigma_3\otimes 1 \right)
        \label{Pitriple}
\end{equation}
and
\begin{equation}
 \Pi_{(d)} = \frac{1}{4}\left(I + 1\otimes1\otimes\sigma_3\otimes\sigma_3
 + 1\otimes\sigma_3\otimes 1\otimes\sigma_3
 +1\otimes\sigma_3\otimes\sigma_3\otimes 1 \right)
        \label{Pisingle}
\end{equation}
are projection operators satisfying the relations
\begin{equation}
   \Pi_{(a)}\Pi_{(b)} = \delta_{ab}\Pi_{(a)},\quad \Pi_{(a)}\pi_j = \delta_{av}\pi_j
        \label{Pirelation}
\end{equation}
for $(a, b = v, d)$ and $(j=1,\cdots,8)$. 

Here we impose a crucial postulate that the operators $\Pi_{(v)}$ and $\Pi_{(d)}$
work to divide the triplet field into the orthogonal component fields as
\begin{equation}
    \Psi_{(v)}(x)=\Pi_{(v)}\Psi(x),\quad \Psi_{(d)}(x)=\Pi_{(d)}\Psi(x)
\end{equation}
which represent, respectively, fundamental fermionic species belonging to the
visible and dark sectors of the Universe. The visible part $\Psi_{(v)}(x)$
can be further decomposed into the sum of the three component fields as follows: 
\begin{equation}
     \Psi_{(v)}(x)=\sum_{j=1,2,3}\Psi_j(x)=\sum_{j=1,2,3}\Pi_j\Psi(x)
\end{equation}
where the projection operators $\Pi_j$ are defined symmetrically by
\begin{equation}
\left\{\ 
 \begin{array}{l}
 \Pi_{1} = \frac{1}{4}\left(I + 1\otimes1\otimes\sigma_3\otimes\sigma_3
 - 1\otimes\sigma_3\otimes 1\otimes\sigma_3
 -1\otimes\sigma_3\otimes\sigma_3\otimes 1 \right),\\
 \noalign{\vskip 0.3cm}
 \Pi_{2} = \frac{1}{4}\left(I - 1\otimes1\otimes\sigma_3\otimes\sigma_3
 + 1\otimes\sigma_3\otimes 1\otimes\sigma_3
 -1\otimes\sigma_3\otimes\sigma_3\otimes 1 \right),\\
 \noalign{\vskip 0.3cm}
 \Pi_{3} = \frac{1}{4}\left(I - 1\otimes1\otimes\sigma_3\otimes\sigma_3
 - 1\otimes\sigma_3\otimes 1\otimes\sigma_3
 +1\otimes\sigma_3\otimes\sigma_3\otimes 1 \right) .
 \end{array}
\right. 
\end{equation}
which obey the relations $\sum_{j=1}^3\Pi_{j}=\Pi_{(v)}$ and $\Pi_j\Pi_k=\delta_{jk}\Pi_j$.
The component fields $\Psi_j(x)\,(j=1,2,3)$ are interpreted to be the fields for three
ordinary families of quarks and leptons in interaction modes.

With the operators $\pi_j$ and $\Pi_{(a)}$, let us construct the set
of the $\mathfrak{su}(3)$ and $\mathfrak{u}(1)$ algebras by
\begin{equation}
  A_{(v)} = \{\ \Pi_{(v)}, \ \pi_1,\,\pi_2, \cdots, \pi_8\ \},\quad
  A_{(d)} = \{\ \Pi_{(d)} \ \}
  \label{Afamily}
\end{equation}
which are closed and irreducible under the action of $S_3$ permutation.
The algebras $A_{(v)}$ and $A_{(d)}$ specify, respectively,
the characteristics of the three ordinary families of the visible sector
and the exotic family of the dark sector. Accordingly, the set
$A_f = \{A_{(v)},\,A_{(d)}\}$ is the algebra of operators specifying
the family structure in the triplet field theory.

Rich varieties observed in flavor physics are presumed in the SM to result
from the Yukawa couplings of quarks and leptons with the Higgs field.
In low energy regime of flavor physics, quarks and leptons manifest
themselves in both of the dual modes of electroweak interaction and
mass eigen-states. It is the elements of the algebra $A_{(v)}$ in Eq.(\ref{Afamily})
that determine the structures of the Yukawa coupling constants which brings about
varieties in the mass spectra and the electroweak mixing matrices. 

It is the algebra $A_{(d)}$ consisting of the single element
$\Pi_4\equiv\Pi_{(d)}$ and the fourth component field $\Psi_4\equiv\Psi_{(d)}$
of the triplet field that determine characteristics of exotic
quarks and leptons belonging to the dark sector of the Universe.
The projection operators $\Pi_j\ (j=1,2,3,4)$ satisfy the relations
$\sum_{j=1}^4\Pi_{j} = I$ and $\Pi_j\Pi_k = \delta_{jk}\Pi_j$.

\section{Algebra for extended color degrees of freedom}
In parallel with the arguments in the preceding section,
it is possible to construct another set of {\lq}{\lq}$\mathfrak{su}(3)$
plus $\mathfrak{u}(1)${\rq}{\rq} subalgebras from the algebra
$A_\rho=\{\,\rho_1,\,\rho_2,\,\rho_3\,\}$ in Eq.(\ref{Arho}). Replacing
$\sigma_a$ with $\rho_a$ in Eq.(\ref{pidef}), we obtain a new set of operators
$\lambda_j\ (j=1, \cdots 8)$ in place of $\pi_j$. Likewise, corresponding to
$\Pi_{(a)}\ (a=v, d)$ in Eqs.(\ref{Pitriple}) and (\ref{Pisingle}), we obtain
operators $\Lambda^{(a)}\ (a=q, \ell)$ which play respective roles to project
out component fields representing the quark-like and lepton-like modes of
the triplet field. 

Then, from Eqs.(\ref{picom}) and (\ref{Pirelation}), we find that the operators
$\lambda_j$ and $\Lambda^{(a)}$ satisfy the relations
\begin{equation}
        [\,\lambda_j,\ \lambda_k\,] = 2f^{(3)}_{jkl}\lambda_l,\quad 
       \{\,\lambda_j,\ \lambda_k\,\} = \frac{4}{3}\delta_{jk}\Lambda^{(q)}
        + 2d^{(3)}_{jkl}\lambda_l
        \label{lambdacom}
\end{equation}
and
\begin{equation}
    \Lambda^{(a)}\Lambda^{(b)} = \delta^{ab}\Lambda^{(a)},\quad
    \Lambda^{(a)}\lambda_j = \delta^{aq}\lambda_j
        \label{Lambdarelation}
\end{equation}
for $(a, b = q, \ell)$ and $(j=1,\cdots,8)$. The operators $\lambda_j$ and
$\Lambda^{(a)}$ enable us to build up the new set of $\mathfrak{su}(3)$
and $\mathfrak{u}(1)$ algebras as follows:
\begin{equation}
  A^{(q)} = \{\ \Lambda^{(q)}, \ \lambda_1,\,\lambda_2, \cdots, \lambda_8\ \}, 
  \quad
  A^{(\ell)} = \{\ \Lambda^{(\ell)} \ \}.
  \label{ellalgebra}
\end{equation} 
It is readily proved that these algebras $A^{(a)}$ satisfy the criterion of $S_3$
irreducibility and are commutative with the algebras $A_\Gamma$, $A_{(v)}$ and
$A_{(d)}$.

The operator for the {\lq\lq}baryon number minus lepton number{\rq\rq} defined by
\begin{equation}\!\!
\begin{array}{lcl}\,
  Q_{B-L}\!&\!\!=&\!\! \frac{1}{3}\Lambda^{(q)}-\Lambda^{(\ell)}\\
   \noalign{\vskip 0.3cm}
  \!&\!\!=\!&\!\!-\frac{1}{3}\left(1\otimes1\otimes\rho_3\otimes\rho_3
 + 1\otimes\rho_3\otimes 1\otimes\rho_3+1\otimes\rho_3\otimes\rho_3\otimes 1\right)
\end{array}
\end{equation}
obeys the minimal equation
\begin{equation}
        \left(Q_{B-L} +I\right)\left(Q_{B-L} - \frac{1}{3}I\right) = 0
\end{equation}
and has the eigenvalues $\frac{1}{3}$ and -1. Therefore, $\Lambda^{(q)}\Psi$ and
$\Lambda^{(\ell)}\Psi$ form, respectively, the quark-like and lepton-like modes
of the triplet field.  

At this stage, we construct the generators for extended color gauge symmetries $SU_c(3)$
and $SU_{c\star}(3)$ which act, respectively, to the visible and dark fields $\Psi_{(v)}$
and $\Psi_{(d)}$. Combining the elements of the core algebras $A^{(q)}$ with
the projection operators $\Pi_{(a)}$, we can make up the operators as 
\begin{equation}
  \Lambda_{(a)}^{(q)} = \Pi_{(a)}\Lambda^{(q)}, \ \lambda_{(a)j}=\Pi_{(a)}\lambda_j
  \label{vdcolor}
\end{equation}
which form the algebras
\begin{equation}
  A_{(a)}^{(q)} = \{\, \Lambda_{(a)}^{(q)}, \ \lambda_{(a)j}: j=1,\,\cdots,8\, \}
  \label{vsu3}
\end{equation}
where $a= v, d$. The elements of the algebras $A_{(a)}^{(q)}$ satisfy the commutation and
anti-commutation relations
\begin{equation}
[\,\lambda_{(a)j},\ \lambda_{(a)k}\,] = 2f^{(3)}_{jkl}\lambda_{(a)l},\ \ 
\{\,\lambda_{(a)j},\ \lambda_{(a)k}\,\}=\frac{4}{3}\delta_{jk}\Lambda_{(a)}^{(q)}
        + 2d^{(3)}_{jkl}\lambda_{(a)l}
\end{equation}
for $a=v,\,d$ and $j,\,k,\,l = 1, 2, \cdots, 8$. 

In this formalism, the quark-like species in the visible and dark sectors are presumed
to be confined separately by different color gauge interactions associated with the
groups $SU_c(3)$ and $SU_{c\star}(3)$. Those gauge groups are defined by the exponential
mappings of the algebras $A^{(q)}_{(a)}\ (a=v,\,d)$ as
\begin{equation}
  SU_c(3)\times SU_{c\star}(3)
  = \left\{ \exp\left(-\frac{i}{2}\sum_{a=v,\,d}\sum_j\lambda_{(a)j}\theta_{(a)}^{j}(x)
     \right)\Lambda^{(q)}\right\}
  \label{colorsymmetries}
\end{equation}
where $\theta_{(a)}^{j}(x)$ are arbitrary real functions of space-time. In addition to
the ordinary gauge fields $A_{\mu}^{(3)j}(x)$ with coupling constant $g^{(3)}$ of the
$SU_c(3)$ symmetry, our theory necessitates the new gauge fields $A_{\star\mu}^{(3)j}(x)$
with coupling constant $g_{\star}^{(3)}$ of the $SU_{c\star}(3)$ symmetry.

For the lepton-like species also, we have to introduce the algebras
\begin{equation}
  A_{(a)}^{(\ell)} = \{\, \Lambda_{(a)}^{(\ell)} \equiv \Pi_{(a)}\Lambda^{\ell} \, \}.
  \label{dsu1}
\end{equation}
with $a= v, d$, which act to the visible and dark sectors, respectively. The set
$A^c=\{\,A^{(q)}_{(v)},\,A^{(\ell)}_{(v)};\,A^{(q)}_{(d)},\,A^{(\ell)}_{(d)} \}$
is the algebra of operators characterizing the extended color degrees of freedom in the
triplet field theory. The operators $Q_{B-L}^{(a)}\ (a=v, d)$ of
{\lq\lq}baryon number minus lepton number{\rq\rq} in the visible and dark sectors
are defined, respectively, by
\begin{equation}
   Q_{B-L}^{(v)}= \Pi_{(v)}Q_{B-L},\ Q_{B-L}^{(d)}= \Pi_{(d)}Q_{B-L}. 
\end{equation}

\section{Fundamental representation of the group $G \times G_\star$} 
The triplet field $\Psi$ with $4\times 4\times 4$ spinor-components is decomposed as
\begin{equation}
 \Psi = \Psi_{(v)} + \Psi_{(d)}
      = \left[\Psi_{(v)}^{(q)}+\Psi_{(v)}^{(\ell)}\right]
      + \left[\Psi_{(d)}^{(q)}+\Psi_{(d)}^{(\ell)}\right]
\end{equation}
where $\Psi_{(a)}^{(c)}=\Lambda^{(c)}\Pi_{(a)}\Psi\ (a=v,\,d; c=q,\,\ell)$
are the four component fields of Dirac-type with degrees of freedom of
$(3+1)$-families and $(3+1)$-colors. Hence the triplet field $\Psi$ which is
considered to be the basic unit of fermionic species has no more freedom.

In order to incorporate the Weinberg-Salam symmetry $G_{EW}$ and its left-right twisted
symmetry $G_{EW\star}$ in the visible and dark sectors, respectively, we have to 
postulate that there exists a two-storied compound field ${\boldsymbol \Psi}$ consisting
of two triplet fields. The compound field ${\boldsymbol \Psi}$ forms the fundamental
representations of the gauge group $G \times G_\star$ as
\begin{equation}
 {\boldsymbol \Psi} = {\boldsymbol \Psi}_L+{\boldsymbol \Psi}_R=
       \left(\, {\boldsymbol \Psi}_{(v)}\ \  
       \begin{array}{ccc}
             U_{(d)}\\
          \noalign{\vskip 0.1cm}
             D_{(d)}\\
       \end{array}
       \right)_L+
       \left( 
       \begin{array}{ccc}
             U_{(v)}\\
          \noalign{\vskip 0.1cm}
             D_{(v)}\\
       \end{array}\ \  
        {\boldsymbol \Psi}_{(d)}\,
     \right)_R
     \label{Fundamentalrep}
\end{equation}
where ${\boldsymbol \Psi}_{(v)L}$, $U_{(v)L}$ and $D_{(v)L}$ (${\boldsymbol \Psi}_{(d)R}$,
$U_{(d)R}$ and $D_{(d)R}$) are, respectively, the chiral multi-spinor fields of the doublet, 
the up singlet and the down singlet of the $SU_L(2)$ ($SU_R(2)$) symmetry. We interpret that
all fermionic species in the visible and dark sectors of the Universe are described by
the component fields of the single compound field ${\boldsymbol \Psi}$.

To name the fermionic species in the dark sector, let us assign new symbols $u_\star$
and $d_\star$ for up and down {\it dark quarks}, and $\nu_\star$ and $e_\star$ for up
and down {\it dark leptons}. Then, the quark parts of the chiral compound fields
${\boldsymbol \Psi}_L$ and ${\boldsymbol \Psi}_R$ are schematically expressed,
respectively, by
\begin{equation}
{\boldsymbol \Psi}^{(q)}_{(v)}
=\left(  
  \begin{array}{ccc}
   u & c & t\\
   d & s & b
  \end{array}
\right)_L,\ 
 U^{(q)}_{(d)}
 =\left( u_\star \right)_L
,\ 
 D^{(q)}_{(d)}
 =\left( d_\star \right)_L
\end{equation}
and
\begin{equation}
U^{(q)}_{(v)}
=\left(  
  \begin{array}{ccc}
    u & c & t
  \end{array}
\right)_R
,\ 
D^{(q)}_{(v)}
=\left(  
  \begin{array}{ccc}
   d & s & b 
  \end{array}
\right)_R
,\ 
{\boldsymbol \Psi}^{(q)}_{(d)}
=\left(  
  \begin{array}{c}
     u_\star\\
     d_\star
  \end{array}
\right)_R.
\end{equation}
Similarly, the lepton parts have the following expressions as
\begin{equation}
{\boldsymbol \Psi}^{(\ell)}_{(v)}
=\left(  
  \begin{array}{ccc}
   \nu_e & \nu_\mu & \nu_\tau \\
   e & \mu & \tau
  \end{array}
\right)_L,\ 
 U^{(\ell)}_{(d)}
 =\left( \nu_\star \right)_L
,\ 
 D^{(\ell)}_{(d)}
 =\left( e_\star \right)_L
\end{equation}
and
\begin{equation}
U^{(\ell)}_{(v)}
=\left(  
  \begin{array}{ccc}
   \nu_e & \nu_\mu & \nu_\tau
  \end{array}
\right)_R
,\ 
D^{(\ell)}_{(v)}
=\left(  
  \begin{array}{ccc}
   e & \mu & \tau 
  \end{array}
\right)_R
,\ 
{\boldsymbol \Psi}^{(\ell)}_{(d)}
=\left(  
  \begin{array}{c}
     \nu_\star\\
     e_\star
  \end{array}
\right)_R.
\end{equation}

The kinetic and gauge parts of the Lagrangian density for all fermions can
now be written down in terms of the chiral compound fields ${\boldsymbol \Psi}_L$
and ${\boldsymbol \Psi}_R$ by
\begin{equation}
 {\cal L}_{kg}  = \bar{{\boldsymbol \Psi}}_L i\Gamma^\mu{\cal D}_\mu
                        {\boldsymbol \Psi}_L
                    + \bar{{\boldsymbol \Psi}}_R i\Gamma^\mu{\cal D}_\mu
                        {\boldsymbol \Psi}_R
\end{equation}
in which the covariant derivatives act as follows:
\begin{equation}
\hspace*{-0.45cm}
\begin{array}{l}
 i{\cal D}_\mu{\boldsymbol \Psi}_L = 
   \left\{i\partial_\mu - \left[g^{(3)}A_{\mu}^{(3)j}\half\lambda_{(v)j}
              + g^{(2)}A_{\mu}^{(2)j}\half\tau_{Lj}
              + g^{(1)}A_{\mu}^{(1)}\half Y\right]\Pi_{(v)} \right.\\
   \noalign{\vskip 0.3cm}
  \hspace*{4.0cm} \left. \ \ -\left[g_\star^{(3)}A_{\star\mu}^{(3)j}\half\lambda_{(d)j} 
+g_\star^{(1)}A_{\star\mu}^{(1)}\half Y_\star\right]\Pi_{(d)}\right\}{\boldsymbol \Psi}_L 
\end{array}
\end{equation}
and
\begin{equation}
\hspace*{-0.25cm}
\begin{array}{l}
 i{\cal D}_\mu{\boldsymbol \Psi}_R = 
  \left\{i\partial_\mu - \left[ g^{(3)}A_{\mu}^{(3)j}\half\lambda_{(v)j}
  + g^{(1)}A_{\mu}^{(1)}\half Y\right]\Pi_{(v)} \right.\\
 \noalign{\vskip 0.3cm}
   \hspace*{1.4cm}  \left. 
  -\left[g_\star^{(3)}A_{\star\mu}^{(3)j}\half\lambda_{(d)j}
  +g_\star^{(2)}A_{\star\mu}^{(2)j}\half\tau_{Rj}
  +g_\star^{(1)}A_{\star\mu}^{(1)}\half Y_\star\right]\Pi_{(d)}\right\}{\boldsymbol \Psi}_R
\end{array}
\end{equation}
where $A_{\mu}^{(2)j}$ and  $A_{\mu}^{(1)}$ ($A_{\star\mu}^{(2)j}$ and $A_{\star\mu}^{(1)}$)
are gauge fields of the $G_{\rm EW}$ ($G_{\rm EW\star}$) symmetry with coupling constants
$g^{(2)}$ and $g^{(1)}$ ($g_\star^{(2)}$ and $g_\star^{(1)}$). The operators $\half\tau_{Lj}$
($\half\tau_{Rj}$) are the generators of the visible (dark) $SU_L(2)$ ($SU_R(2)$) symmetry,
and the hypercharge $Y$ ($Y_\star$) in the visible (dark) sector can be expressed by
\begin{equation}
      Y = Q_{B-L}^{(v)} + y,\quad Y_\star = Q_{B-L}^{(d)} + y_\star
\end{equation}
in which $y$ ($y_\star$) takes 0, 1 and -1 for the doublet $\Psi$, the up singlet $U$
and the down singlet $D$. 

To break down the gauge symmetries $G_{\rm EW}$ and $G_{\rm EW\star}$, we require two types
of Higgs doublets $\varphi$ and $\varphi_\star$ which, respectively, have the hypercharges
($Y=1$,\, $Y_\star=0$) and ($Y=0$,\, $Y_\star=1$). The Lagrangian density of the Yukawa
interaction is given as follows:
\begin{equation}
 \begin{array}{l}
  {\cal  L}_Y =          
    \bar{{\boldsymbol \Psi}}_L\left\{\,{\rm Higgs\ fields}\,\right\}{\boldsymbol \Psi}_R
       +  {\rm h.c.}\\
      \noalign{\vskip 0.4cm}
      \hspace*{0.64cm}
       = \bar{{\boldsymbol \Psi}}_{(v)}\tilde{\varphi}{\cal Y}_{U}U_{(v)}
         +  \bar{{\boldsymbol \Psi}}_{(v)}\varphi{\cal Y}_{D}D_{(v)}
         +  y_{u \star}\bar{U}_{(d)}\tilde{\varphi}_\star^\dagger {\boldsymbol \Psi}_{(d)}
         +  y_{d \star}\bar{D}_{(d)}\varphi_\star^\dagger {\boldsymbol \Psi}_{(d)} + {\rm h. c.}
 \end{array}
 \label{Yukawa}
\end{equation}
where $\tilde{\varphi}=i\tau_{L2}\varphi^\ast$ and
$\tilde{\varphi}_\star=i\tau_{R2}\varphi_\star^\ast$.
The operators ${\cal Y}_{U}$ and ${\cal Y}_{D}$ consisting of elements of the algebra
$A_{(v)}$ in Eq.(\ref{Afamily}) determine the patterns of Yukawa interactions among
the fermions in the visible sector~\cite{Sogami2}, and $y_{u \star}$ and $y_{d \star}$
are the Yukawa coupling constants of the fermions in the dark sector. 

The Lagrangian density of the visible and dark Higgs fields takes the form
\begin{equation}
    {\cal L}_{H}= ({  D}^\mu\varphi)^\dagger({  D}_\mu\varphi)
                + ({  D}^\mu\varphi_\star)^\dagger({  D}_\mu\varphi_\star)
                - V_H 
    \label{Higgs}    
\end{equation}
in which the covariant derivatives act as follows:
\begin{equation}
 i{  D}_\mu \varphi = 
              \left(i\partial_\mu - g^{(2)}A_{\mu}^{(2)a}\half\tau_{La}
              - g^{(1)}A_{\mu}^{(1)}\half \right)\varphi  
\end{equation}
and
\begin{equation}\ \ \ 
 i{  D}_\mu \varphi_\star = 
              \left(i\partial_\mu - g_\star^{(2)}A_{\star\mu}^{(2)a}\half\tau_{Ra}
              - g_\star^{(1)}A_{\star\mu}^{(1)}\half \right)\varphi_\star .  
\end{equation}
The Higgs potential is generally given by
\begin{equation}
    V_H = V_0 - \mu^2 \varphi^\dagger\varphi + \lambda(\varphi^\dagger\varphi)^2
        - \mu_\star^2 \varphi_\star^\dagger\varphi_\star
        + \lambda_\star(\varphi_\star^\dagger\varphi_\star)^2
        + 2\lambda_I (\varphi_\star^\dagger\varphi_\star)(\varphi^\dagger\varphi)
    \label{HiggsPot}                
\end{equation}
where $\lambda,\,\lambda_\star$ and $\lambda_I$ are the constants of self-coupling
and bi-quadratic mutual interaction.

\section{A scenario for dark matter\label{scenarios}}
If the quarks $u_\star$ and $d_\star$ acquire close masses like the $u$ and $d$
quarks of the first family when the $G_{EW\star}$ symmetry is spontaneously broken
at the energy scale $\Lambda_\star$ ($\Lambda_\star > \Lambda$), many kinds of dark
nuclei and a variety of dark elements come necessarily into existence. Thereby,
the dark sector with such quarks $u_\star$ and $d_\star$ is destined to follow a rich
and intricate path of thermal history of evolution like the visible sector of our
Universe.

Here we consider a situation that, just like the case of the $t$ and $b$ quarks
of the third family, the dark up quark $u_\star$ is much heavier than the dark
down quark $d_\star$ as~\cite{Sogami1,Sogami3}
\begin{equation}
   m_{u_\star} \gg m_{d_\star} + m_{e_\star} + m_{\nu_\star}.
\end{equation}
In such a case, the $u_\star$ quark disappears quickly through the process
$u_\star \rightarrow d_\star + \bar{e}_\star + \nu_\star$ leaving the $d_\star$ quark
as the main massive components of the dark sector. Consequently, the gauge fields
$A_{\star\mu}^{(3)}$ of $SU_{c\star}(3)$ symmetry act to confine the $d_\star$ quark
into the dark color-singlet hadron
\begin{equation}
      \Delta^-_\star = [d_\star\,d_\star\,d_\star] 
      = \frac{1}{\sqrt{6}}\epsilon_{ijk}d^i_\star\,d^j_\star\,d^k_\star   
\end{equation}
which has the dark electric charge $Q_\star = -1$ and the spin angular momentum $\frac{3}{2}$
due to the Fermi statistics.

In this scenario, $\Delta^-_\star$ is the stable dark hadron which can exist as the only
nucleus in the dark sector. Therefore, no rich nuclear reaction can occur and only a meager
history of thermal evolution can take place. The stable atom which can exist in
the matter-dominant stage of the dark sector is limited to be
\begin{equation}
        \bar{H}_\star = (\Delta_\star^- + \bar{e}_\star).
\end{equation}
It can be speculated that the dark molecule
\begin{equation}
        (\bar{H}_\star)_2 = \bar{H}_\star \bar{H}_\star 
\end{equation}
in the spin 1 or 0 states can be stable entities prevailing over a broad spatial region
of the late stage of the Universe. These features seem to be consistent with the
characteristics of the dark matter which has the tendency to spread out rather
monotonically over broad spatial regions as inferred by the observations of
gravitational lensing.

\section{Discussion}
By generalizing the concept of Dirac spinor, we have developed a unified theory of
multi-spinor fields that can describe the whole spectra of fermionic species in
the visible and dark sectors of the Universe. The physical subalgebras of the triplet
algebra satisfying the criterion of the $S_3$ irreducibility have the unique feature
of the {\lq}{\lq}$\mathfrak{su}(3)$ plus $\mathfrak{u}(1)${\rq}{\rq} structure
for both of the color and family degrees of freedom. The triplet compound field
in Eq.(\ref{Fundamentalrep}) possessing the component fields of the three visible
and one dark family modes with the tri-color quarks and colorless leptons enables
us to formulate a unified theory that can describe the flavor physics in
the visible sector and cosmological phenomena related to both the visible
and dark sectors. 

To develop the present theory further, it is necessary to examine the thermal history
of the dark sector and to confirm that the present scheme is consistent with the
well-established standard theory of astrophysics and cosmology. In this note, we make
heuristic analyses to estimate the mass of the stable dark hadron $\Delta_\star$ from
the cosmological parameters for the densities of the cold dark matter and baryonic
matter determined by WMAP~\cite{WMAP} and Planck~\cite{Planck} observations.  

The visible fermionic and bosonic fields can interact with the dark fermionic
and bosonic fields through virtual loop corrections induced by the bi-quadratic
interaction between the Higgs fields $\varphi$ and $\varphi_\star$ in
Eq.(\ref{HiggsPot}). Therefore, it is possible to assume that quanta of
all fields of the visible and dark sectors constitute a common soup of
inseparable phase in an early reheating period. Expansion of the Universe
decreasing its temperature breaks both of the dark electroweak symmetry
$G_{EW\star}$ and the visible electroweak symmetry $G_{EW}$ symmetry in the
quantum soup. 

At present, there exists no reliable theory which can describe consistently the
cosmic baryogenesis. So we set a simple working hypothesis~\footnote{Technical
details of some scenarios of {\lq}Two-Step Electroweak Baryogenesis{\rq} can be
found in the articles~\cite{BG1,BG2}.} that the process of baryogenesis takes
place cooperatively through the two-step breakdowns of $G_{EW\star}$ and $G_{EW}$
symmetries in such a way that the excess of quark numbers is created and preserved equally for all families in the visible and dark sectors. The quarks getting heavy
masses decay down to the lighter quarks. While the $u_\star$ quark disappears
leaving only the $d_\star$ quark in the dark sector, four types of the heavy
quarks decay into the $u$ and $d$ quarks which have almost degenerate masses
in the visible sector. The $SU_{c\star}(3)$ gauge interaction works to confine
the $d_\star$ quarks into the dark hadron $\Delta_\star$ with 4 spin degrees
of freedom, and the $SU_c(3)$ gauge interaction confine the $u$ and $d$ quarks
into the nucleon possessing 2 iso-spin (proton and neutron) states with 2 spin
degrees of freedom. 

Remark that the quark numbers are separately conserved in the dark and visible sectors
after the decoupling of the two sectors. Therefore, the observed ratio of the densities
of cold dark matter and baryonic matter can be identified with the ratio of energies
(masses/c$^2$) stored by the stable particles in the dark and visible sectors.
By assuming that the dark hadron $\Delta_\star$ is the dominant component of
the cold dark matter, we obtain the following relation for the masses of
$\Delta_\star$ and nucleon, $m_{\Delta_\star}$ and $m_N$, as
\begin{equation}
   2m_{\Delta_\star} : 6m_N = \Omega_c h^2 : \Omega_b h^2 = 0.11889 : 0.022161
\end{equation}
where the values for the cosmological parameters $\Omega_c h^2$ and $\Omega_b h^2$
taken from the Table 10 of the reference~\cite{Planck} are the Planck best-fit
including external data set. Consequently, the upper limit of the mass of the dark hadron
$\Delta_\star$ is estimated to be $m_{\Delta_\star} = 16.1 m_N = 15.1$GeV/${\rm c}^2$.

Until now no affirmative result has been found by either of the direct and indirect
dark matter searches. The recent observation by the LUX group~\cite{LUX} has proved that
the background-only hypothesis is consistent with their data on spin-independent
WIMP-nucleon elastic scattering with a minimum upper limit on the cross section
of $7.6\times 10^{-46}$cm$^2$ at a WIMP mass of 33 GeV/${\rm c}^2$. More stringent
experiments must be performed to confirm the possibility of scenarios for dark matter
including stable particles with comparatively small masses such as the dark
hadron $\Delta_\star$ of our theory.   




\begin{thebibliography}{99}
\bibitem{Sogami1} I.~S.~Sogami, Prog. Theor. Exp. Phys. \textbf{123} B02 (2013). 
\bibitem{Sogami2} I.~S.~Sogami, Prog. Theor. Phys. \textbf{78} 487 (1987);
                  J. Phys.: Conf. Ser. \textbf{343}, 012113 (2012).
\bibitem{Sogami3} I.~S.~Sogami, JPS Conf. Proc. \textbf{7}, 010005 (2015).
\bibitem{WMAP} E. Komatsu et al. (WMAP Collaboration), Astrophys. J. Suppl.
\textbf{192}, 18 (2011), 1001.4538.
\bibitem{Planck} P.~A.~R.~Ade et al. (Planck Collaboration), Astron.Astrophys.
\textbf{571}, A16 (2014) 1303.5076.
\bibitem{BG1} D.~Land and E.~D.~Carlson, Phys. Lett. \textbf{B292}, 107 (1992). 
\bibitem{BG2} S.~Inoue, G.~Ovanesyan and M.~J.~Ramsey-Musolf, arXiv:[1508.0540v1]
[hep-ph], (Accepted in Phys. Rev. D), and references therein. 
\bibitem{LUX} D.~S.~Akerib et al. (LUX Collaboration), Phys. Rev. Lett. \textbf{112},
091303 (2014).
\end{thebibliography}
\end{document}